\documentclass[twocolumn,showkeys,showpacs,preprintnumbers,prd,superscriptaddress,nofootinbib,aps,10pt]{revtex4-1}
\usepackage{graphicx,epsf,bm,amsmath,amsfonts,amssymb,epstopdf,natbib,hyperref,color,verbatim,multirow,bm,appendix}
\hypersetup{colorlinks=true,urlcolor=blue,citecolor=blue,linkcolor=blue,menucolor=blue,anchorcolor=blue,filecolor=blue}
\usepackage{bbm}
\begin{document}
\title{A Fast Method to Compute Scalar Induced Gravitational Waves on a Lattice with Primordial  Non-Gaussianities}

\author{Giovanni Piccoli}
\email{giovanni.piccoli@uzh.ch}

\affiliation{%
 University of Zurich, Department of Astrophysics\\
}%
\date{\today}

\begin{abstract}
Scalar Induced Gravitational Waves (SIGW) are generated at second order in perturbation theory and to achieve observational relevance, inflationary dynamics must evade the standard slow-roll scenario at small scales, generating large curvature perturbations following strongly non-Gaussian statistics. We propose a method to efficiently compute the SIGW spectrum including arbitrary non-Gaussianities. First, we solve the wave equation adopting semi-analytic methods; this results in an expression involving integrals in Fourier space which are impossible to solve directly on a lattice. We overcome this bottleneck by recasting these integrals
as a sum of $ \sim 50$ convolutions, each of which can be computed efficiently with FFT methods. Finally, the power spectrum is measured directly from the lattice realization. We implement this in \texttt{FLAN-SIGW}, a GPU-accelerated code capable of computing fully non-perturbative, non-Gaussian SIGW spectra in seconds with an error within $\sim 10\%$ with modest computational resources. The code is made public on \href{https://github.com/giovannipiccoli99/FLAN-SIGW}{GitHub}.  In this first implementation, in order to assess the performance of the method, we adopt a standard radiation-dominated background with $w = 1/3$. 
\end{abstract}

\maketitle


Gravitational wave astronomy is a promising avenue to obtain information about the evolution of the earliest phase of the Universe. 
During a period of cosmic Inflation \cite{Guth:1980zm}, quantum fluctuations would be responsible not only for generating the seeds of the Large Scale Structure \cite{Starobinsky:1980te, Mukhanov:1981xt}, but also primordial gravitational waves. The amplitude of the primordial curvature power spectrum is constrained to $\Delta_\zeta^2 \sim 10^{-9}$ on large scales ($k\sim 0.05$ Mpc$^{-1}$), while its spectral tilt is slightly red ($n_s \lesssim 1$) \cite{AtacamaCosmologyTelescope:2025nti}; these highly Gaussian fluctuations align precisely with standard single-field slow-roll predictions. 
Due to the non-linear nature of Einstein's equations, a quadratic combination of first-order scalar perturbations sources second-order tensor modes \cite{Tomita:1967wkp,Matarrese:1992rp,Matarrese:1993zf, Matarrese:1997ay, Ananda:2006af, Baumann:2007zm}, known as Scalar Induced Gravitational Waves (SIGW). For a detailed review, we refer to \cite{Domenech:2021ztg}.  Primary tensor modes, which would imprint in the B-modes of the  polarization of the Cosmic Microwave Background, remain undetected \cite{Tristram:2020wbi}, and the secondary SIGW contribution on these cosmological scales is subdominant; moreover, an extrapolation of the slow-roll power spectrum to smaller scales yields a negligible SIGW amplitude. However, while the scalar power spectrum is well known on large scales ($k\lesssim 1$ Mpc$^{-1}$), on smaller scales it remains largely unconstrained, apart from upper bounds coming from the non-observations of Spectral Distortions \cite{Chluba:2012we, Nakama:2017ohe, Chluba:2014sma}, Acoustic Reheating \cite{Jeong:2014gna,Naruko:2015pva,Inomata:2016uip,PhysRevLett.113.061302, Piccoli:2026grg, Ota:2017jte}, Primordial Black Holes \cite{Josan:2009qn} and signatures of Ultra-Compact Mini-Halos \cite{Bringmann:2011ut}. A transient deviation from the slow-roll attractor such as a phase of ultra-slow-roll \cite{Dimopoulos:2017ged} could lead to a large enhancement in the scalar power spectrum at scales which are observationally reachable as the nanohertz band, probed by Pulsar Timing Arrays (PTA) \cite{NANOGrav:2023hvm, EPTA:2023xxk, Ellis:2023oxs, Balaji:2023ehk, Cecchini:2025oks}, or the millihertz band, probed by the Laser Interferometer Space Antenna (LISA) \cite{LISACosmologyWorkingGroup:2022jok, LISACosmologyWorkingGroup:2025vdz}. Interestingly, a deviation from the slow-roll attractor not only translates into a boost in the amplitude of the perturbation, but also into a deviation from the approximate Gaussianity typical of perturbations produced during a slow-roll phase. Due to their non-linear origin, the two-point function of the SIGW is given in terms of the four-point function of the scalar modes, $\langle hh\rangle \supset\langle \zeta\zeta\zeta\zeta\rangle$; the computation of this four-point function adopting Wick's theorem \cite{PhysRev.80.268} has been performed semi-analytically assuming the non-Gaussianities to be amenable to a local, perturbative treatment \cite{Cai:2018dig, Perna:2024ehx}. An interesting possibility is that such a perturbative expansion may not be possible \cite{Iovino:2024sgs}; under such circumstance, a full numerical simulation would be needed \cite{Zeng:2025cer, Caravano:2026hca}. Moreover, there is evidence for the failure of the leading order separate universe approximation \cite{Jackson:2023obv} for sudden transitions to a non-attractor phase, which points to the requirement of laplacian corrections, jeopardizing the validity of the local expansion. Numerical simulations have also been employed for mixed adiabatic-isocurvature initial conditions \cite{Zeng:2025tno}. A measurement of the power spectrum from a lattice realization of a given field is completely agnostic of the underlying statistics; the computational overload is therefore the same as in the Gaussian case, as opposed to semi-analytic methods, where the number of n-point functions needed to account for local type non-Gaussianities and the dimensionality of the correspondent momentum integrals grows very quickly. In this work, we present a novel method with which the induced gravitational wave spectrum $\Omega_{\rm gw}(k)$ can be computed for a given realization of the primordial comoving curvature $\zeta$. Our approach amounts to two steps. First, the oscillating part of the strain is separated from its slowly varying amplitude, whose asymptotic value can be obtained via a direct integration in conformal time. This leads to expressions for the amplitudes already known from analytic studies in the Gaussian case. While formally exact and amenable to be simplified taking Wick's contractions (when the statistical properties of the scalar modes make it possible), the resulting expressions are numerically intractable, being integrals in Fourier space which cannot be recast as convolutions. The second step, which is the truly new contribution of this work, consists of identifying a convenient way to decompose the integrands so to be able to apply the convolution theorem term by term. Once a model for the non-Gaussian scalar perturbation is specified, this approach allows for very efficient simulations of the induced gravitational wave background, from which the power spectrum can be directly measured with no assumption whatsoever on the underlying statistics. 

The paper is organized as follows. In Sec. \ref{sec_sigw} we briefly summarize the basics of SIGW, mostly for the purpose of establishing the notation. In Sec. \ref{sec_envelope} we derive the equations satisfied by the slowly-varying amplitudes, while in Sec. \ref{sec_envelope_obs} we show how to connect their power spectra to the quantity of observable interest, $\Omega_{\rm gw}$. In Sec. \ref{sec_lattice} we present the detailed algorithm with which the SIGW spectrum can be computed assuming that the decomposition of the integrand exists, while in Sec. \ref{sec_decomposition} we show explicitly how to build said decomposition. Finally, in Sec. \ref{sec_numerical} we benchmark the validity of our method in the simplest Gaussian case, comparing the resulting spectrum with the semi-analytic result, and in Sec. \ref{sec_conclusions} we draw our conclusions, remarking the limitations of this work and establishing future directions.
\section{Scalar Induced Gravitational Waves}
\label{sec_sigw}
In this section we review the standard formalism of SIGW for the sake of establishing the notation and paving the way to the subsequent numerical implementations. Neglecting first order tensors, we write the metric in the Poisson Gauge as follows:
\begin{equation}
    ds^2 = a^2(\eta) \Big[-(1 + 2\Phi)d\eta^2 + \Big[(1 + 2\Psi) \delta_{ij} +  h_{ij}\Big]dx^\alpha dx^\beta\Big],\nonumber
\end{equation}
where $\eta$ is the conformal time and $a(\eta)$ the scale-factor. $h_{ij}$ is a transverse ($\partial^ih_{ij} = 0$), trace-less ($h^i_i = 0$) tensor which we consider to contain only the second-order gravitational waves sourced by the scalar modes.
We focus on such gauge for simplicity and concreteness, minding the possibility of gauge ambiguities, which are however unimportant in the sub-Horizon regime \cite{Kugarajh:2025pjl}.
We further neglect scalar anisotropic stress, so that $\Psi = -\Phi$ \cite{Dodelson:2020bqr}; to second order in perturbation theory, extracting the transverse traceless components of Einstein equations gives the wave equation governing the evolution of the second-order tensors:
\begin{equation}
     h''_{ij} + 2\mathcal{H}h'_{ij} -\nabla^2 h_{ij} = -4\hat{\mathcal{T}}^{lm}_{ij}S_{lm}.
     \label{eq_wave_tensor}
\end{equation}
the un-projected source reads, to second order \cite{Ananda:2006af, Baumann:2007zm}:
\begin{align}
    &S_{ij} =\partial_i\Phi \partial_j\Phi +\frac{2}{3(1 + w)}\partial_i(\Phi + \mathcal{H}^{-1}\Phi')\partial_j(\Phi + \mathcal{H}^{-1}\Phi'),
    \label{eq_source}
\end{align}
where $\mathcal{H} = aH$ is the conformal Hubble factor, $w$ is the equation of state of the background, and we neglected a total divergence $\sim \partial_i(\Phi\partial_j \Phi)$ which does not survive the transverse-traceless projector $\hat{\mathcal{T}}^{lm}_{ij}$, a non-local operator given in Fourier space as follows:
\begin{equation}
[\hat{\mathcal{T}}^{lm}_{ij}S_{lm}]_{\bm k} = \sum_\lambda e^\lambda_{ij}(\hat{\bm k})e_{\lambda}^{lm}(\hat{\bm k})S_{lm}(\bm{k}).
\end{equation}
 The polarization tensors $\bm e^\lambda(\hat{\bm k})$ can be written in terms of two unit vectors $\hat{\bm u}$, $\hat{\bm v}$ forming an orthonormal basis together with $\hat{\bm k}$: 
\begin{align}
        \bm e^{+}(\hat{\bm k}) &= \frac{1}{\sqrt{2}}\Big[\hat{\bm u}\otimes \hat{\bm u} -\hat{\bm v}\otimes \hat{\bm v}\Big], \nonumber\\\quad \bm e^{\times}(\hat{\bm k}) &= \frac{1}{\sqrt{2}}\Big[\hat{\bm u}\otimes \hat{\bm v} +\hat{\bm v}\otimes \hat{\bm u}\Big];
\end{align}
for each wavenumber ${\bm k}$, the polarization tensors can be build first by picking a unit vector $\hat{\bm u}$ orthogonal to $\hat{\bm k}$ (for instance, by considering $\hat{\bm u} = \hat{\bm k}\times \hat{\bm x}/|\hat{\bm k}\times \hat{\bm x}|$ for $\hat{\bm k}\neq \hat{\bm x}$), and then forming the third one by means of the cross product $\hat{\bm v} = \hat{\bm k}\times \hat{\bm u}$. We remark that the map $\hat{\bm k}\to \hat{\bm u}$ is necessarily discontinuous because of the impossibility to define a continuous, non-vanishing tangent vector field on a sphere\footnote{A topological obstruction colloquially known as \textit{hairy-ball theorem}.}.
Finally, to linear order, the potential $\Phi$ evolves according to the following equation \cite{Mukhanov:2005sc}:
\begin{equation}
        \Phi'' + 3\mathcal{H}(1 + c_s^2)\Phi' + (2 \mathcal{H}' + (1 + 3c_s^2)\mathcal{H}^2 - c_s^2 \nabla^2)\Phi= 0,
        \label{eq_scalar}
\end{equation}
where the sound speed reads $c_s^2 = P'/\rho'$. In Fourier space, this equation can be solved in terms of a transfer function $\Phi_{\bm k}(\eta) = T(\eta, k)\Phi_{\bm k, i}$, where the initial condition $\Phi_{\bm k, i}$ is linked to the comoving primordial curvature perturbation $\zeta$ as follows:
\begin{equation}
    \Phi_{\bm k, i} = \frac{3(1  + w)}{5 + 3w}\zeta_{\bm k},
\end{equation}
where $w$ has to be evaluated at Horizon re-entry. The transfer function $T$ is the solution to Eq. \ref{eq_scalar} substituting $\nabla^2\to -k^2$, subjected to the initial conditions $T = 1$, $T' = 0$. 
Moving to Fourier space, we can write the wave equation for each polarization mode: 
\begin{equation}
    h''_{\lambda, \bm k} + 2\mathcal{H}h'_{\lambda, \bm k} + k^2 h_{\lambda, \bm k} = \mathcal{S}_{\lambda, \bm k}(\eta),
    \label{eq_wave_k}
\end{equation}
in terms of the source function $\mathcal \mathcal{S}_{\lambda, \bm k}$ (to lighten the notation, from now on we will write the initial conditions for the potential simply as $\Phi_{\bm q}$):
\begin{align}
   & \mathcal{S}_{\lambda, \bm k}(\eta)   = 4e^{ij}_{\lambda}(\hat{\bm k})\int\frac{d^3\bm q_1}{(2\pi)^3}\int\frac{d^3\bm q_2}{(2\pi)^3}\times\\
   \times (2\pi)^3 &\delta_D^{(3)}(\bm q_1 + \bm q_2 - \bm k)q_{1,i} q_{2, j}\Phi_{\bm q_1}\Phi_{\bm q_2}f(\eta, q_1, q_2) \nonumber.
\end{align}
The kernel function $f$ reads:
\begin{align}
    &f(\eta, q_1, q_2) = T(\eta, q_1) T(\eta, q_1) +\nonumber  \\+& \frac{2}{3(1 + w)}\Big(T + \mathcal{H}^{-1} \partial_\eta T\Big)_{q_1}\Big(T +  \mathcal{H}^{-1} \partial_\eta T\Big)_{q_2}.
\end{align}
From Eq. \ref{eq_wave_k} it is clear that apart from the non-trivial evolution induced from the source term, the motion of each mode has both an oscillating and decaying component, respectively induced by $k^2 h_{\lambda, \bm k}$ and by $2\mathcal{H}h'_{\lambda, \bm k}$. In the following section we will proceed to factor out these components, so to be able to focus on the slowly varying amplitudes, whose evolution carries the non-trivial imprint of the source term. This is the preliminary step of our numerical implementation.
\section{Envelope Equations}
\label{sec_envelope}
The scalars sourcing Eq. \ref{eq_wave_k} decay rapidly as $\Phi_{\bm k}\sim 1/(k\eta)^2$ shortly after horizon crossing. For this reason, for $\eta \gg 1/k$, the induced tensor modes behave as free waves, since $\mathcal{S}_{\lambda, \bm k}\approx 0$. Resolving these fast, free oscillations via a brute-force numerical integration on a 3D lattice is computationally heavy and yields little observational information, given that fast oscillations in $k$ space have to be averaged out anyway (see Sec. \ref{sec_envelope_obs}). Therefore, to efficiently solve Eq. \ref{eq_wave_k}, the first step is to separate the highly oscillatory behavior of each wave from the non-trivial evolution of its amplitude, which is induced while the source is active. To do so, we adopt the method of variation of constants \cite{coddington1955theory} (equivalently, the Green's function method). We first consider the two independent solutions to the homogeneous wave equation, $y_n(k, \eta)$:
\begin{equation}
    y_n'' + 2\mathcal{H}y'_n + k^2 y_n = 0, \quad n = 1, 2.
\end{equation}
These solutions can be obtained numerically for any given thermal history encoded in the comoving Hubble factor $\mathcal{H}$. By factoring out these homogeneous solutions, the computational burden is drastically reduced: numerical integration is only required once per wavenumber magnitude $k$, completely bypassing the need to explicitly compute every orientation of the wave-vector $\bm k$ on the lattice.

With these functions, for each polarization mode $\lambda$ the strain can be written by introducing two varying envelope functions $A$, $B$: 
\begin{equation}
    h_{\lambda, \bm k}(\eta) =A_{\lambda, \bm k}(\eta) y_1(k, \eta) + B_{\lambda, \bm k}(\eta) y_2(k,\eta).
    \label{eq_envelope_ansatz}
\end{equation}
At this point, the number of effective degrees of freedom has been doubled; it is therefore necessary to add a constraint relating the two amplitudes, so to remove the spurious degree of freedom. Given the freedom we have in doing so, we can choose the most convenient:
\begin{equation}
       A'_{\lambda, \bm k} y_1 +  B'_{\lambda, \bm k} y_2 = 0.
       \label{eq_envelope_constraint}
\end{equation}
We can insert the ansatz of Eq. \ref{eq_envelope_ansatz} in Eq. \ref{eq_wave_k}; using Eq. \ref{eq_envelope_constraint}, we obtain the following dynamical equation:
\begin{equation}
    A'_{\lambda, \bm k} y_1' + B'_{\lambda, \bm k} y_2' = \mathcal{S}_{\lambda, \bm k}.
\end{equation}
This last result can be combined with Eq. \ref{eq_envelope_constraint} to disentangle the evolution of each amplitude:
\begin{equation}
    A'_{\lambda, \bm k}  = -  \frac{y_2 \mathcal{S}_{\lambda, \bm k} }{W(\eta, k)}, \quad B'_{\lambda, \bm k} = \frac{y_1 \mathcal{S}_{\lambda, \bm k} }{W(\eta, k)},
\end{equation}
written in terms of the Wronskian determinant: 
\begin{equation}
    W(\eta, k) := y_1(\eta, k)y_2'(\eta, k) - y_1'(\eta, k) y_2(\eta, k).
\end{equation}
Using the homogeneous wave equation it is straightforward to see that the Wronskian satisfies the equation $W' = - 2\mathcal{H}W$, which can be readily solved: 
\begin{equation}
    W(\eta, k) = W_i(k)\Big(\frac{a(\eta)}{a_i}\Big)^{-2},
\end{equation}
a result known as Abel's identity.
We therefore obtain the final form of the equations governing the evolution of the envelopes: 
\begin{align}
    A'_{\lambda, \bm k} &= -  W_i^{-1}y_2(k, \eta) S_{\lambda,\bm k}(\eta) \Big(\frac{a(\eta)}{a_i}\Big)^{2},\nonumber\\ \quad B'_{\lambda, \bm k} &=   W_i^{-1}y_1(k, \eta) S_{\lambda,\bm k}(\eta) \Big(\frac{a(\eta)}{a_i}\Big)^{2}. \label{eq_envelopes}
\end{align}
These equations are exact, and valid for a generic expansion history. We remark that modes of interest to PTA and laser interferometers such as LISA entered the Horizon deep during radiation domination, respectively across the QCD \cite{Franciolini:2023wjm} and Electroweak \cite{Escriva:2024ivo} crossovers. During phase transitions the equation of state decreases from the fiducial value $w < 1/3$, leading to a softer pressure and decreased suppression of perturbations, ultimately resulting in a boost in the amplitude of the scalar induced gravitational waves. However, to cleanly isolate the performance of the proposed lattice algorithm from the effects of realistic expansion histories, we fix $w = 1/3$ in this work. This choice provides a mathematically tractable baseline that is firmly established in the literature and routinely used to confront current or upcoming observational data \cite{NANOGrav:2023hvm, Ellis:2023oxs, Ellis:2023oxs}. The integration of more realistic thermal histories is deferred to future work. Within this choice, the transfer function is simply given in terms of the spherical Bessel function of first order:
\begin{equation}
       T(\eta, k ) =  \frac{3j_1(k\eta/\sqrt{3})}{k\eta/\sqrt{3}}.
\end{equation}
Moreover, the solutions to the homogeneous wave equation are simply given by:
\begin{equation}
    y_1(k, \eta) = \frac{\cos k\eta}{k\eta}, \quad y_2(k, \eta) = \frac{\sin k\eta}{k\eta}, 
\end{equation}
while $W(\eta, k) = 1/k\eta^2$, and $a(\eta) /a_i = \eta$. Therefore, the envelope equations simplify to the following form: 
\begin{align}
    A'_{\lambda, \bm k} &= -  \eta \sin(k\eta) S_{\lambda,\bm k}(\eta),\nonumber\\ \quad B'_{\lambda, \bm k} &=   \eta \cos(k \eta) S_{\lambda,\bm k}(\eta) . 
    \label{eq_envelopes_rd}
\end{align}
Before presenting our novel approach for the solution of Eqs. \ref{eq_envelopes_rd}, we briefly comment on how to connect the envelope amplitudes to observables.
\section{Connecting the Envelopes to Observations}
\label{sec_envelope_obs}
A crucial quantity adopted to characterize stochastic backgrounds is the gravitational wave spectrum $\Omega_{\rm gw}$, defined as the amount of energy density carried by the modes in a given logarithmic bin compared to the critical density of the Universe: 
\begin{equation}
    \Omega_{\rm gw}(k) := \frac{1}{\rho_c}\frac{d\rho_{\rm gw}}{d\log k}. 
\end{equation}
A standard computation links the expected value of the gravitational wave spectrum to the dimensionless power spectra of each polarization $\Delta^2_{h, \lambda}$:
\begin{equation}
    \Omega_{\rm gw}(\eta, k) = \frac{k^2}{12\mathcal{H}^2}\sum_{\lambda = +, \times}\Delta_{h, \lambda}^2(\eta, k),
\end{equation}
We remark that for a stochastic field $X$, $\Delta^2_X$ is connected to the dimensionful power spectrum $P_X$ as follows:
\begin{equation}
    \Delta^2_X(k) := \frac{k^3}{2\pi^2}P_X(k),
    \label{eq_adimensional_P}
\end{equation}
while $P_X$ is defined in terms of the two-point function in Fourier space: 
\begin{equation}
    \langle X^*(\bm k')X(\bm k)\rangle = (2\pi)^3 \delta^{(3)}_D(\bm k - \bm k') P_X(k). 
\end{equation}
For concreteness, we now consider the case of pure radiation domination, where Eq. \ref{eq_envelope_ansatz} becomes:
\begin{equation}
    h_{\lambda, \bm k}(\eta) =\frac{1}{k\eta}\Big[A_{\lambda, \bm k}(\eta) \cos k\eta + B_{\lambda, \bm k}(\eta) \sin k\eta\Big];
\end{equation}
therefore, one has, at a given time $\eta$:
\begin{align}
   &P_{h, \lambda}(\eta, k)  = \frac{1}{(k\eta)^2}\Big[P_{A, \lambda}(\eta, k) \cos^2 k\eta + \\ +  &P_{B, \lambda} (\eta, k) \sin^2 k\eta + P_{AB, \lambda}(\eta, k) \sin(2k\eta)\Big];\nonumber
\end{align}
using Eq. \ref{eq_adimensional_P} and using the fact that in radiation era $\mathcal{H} = 1/\eta$, we can then write for $\eta \gg 1/k$:
\begin{align}
    &\Omega_{\rm gw}(\eta, k) = \frac{k^3}{24\pi^2}\sum_{\lambda = +, \times}\Big[P_{A, \lambda}(k) \cos^2 k\eta + \\ &+  P_{B, \lambda} ( k) \sin^2 k\eta + P_{AB, \lambda}( k) \sin(2k\eta)\Big];\nonumber
\end{align}
where now we consider the asymptotic values of $P_{A}, P_{B}, P_{AB}$, considering that for $\eta \gg 1/k$ the envelopes become constant. Furthermore, we may consider that observations are not able to track fast oscillations taking place in $k$ space: as customary, we take the constant mode of the oscillating factors, which for $\cos^2k\eta$ and $\sin^2k\eta$ is $1/2$ and for $\sin2k\eta$ is $0$, which finally leaves us with the asymptotic value of the gravitational wave spectrum deep in radiation domination:
\begin{align}
    \Omega^{(\rm RD)}_{\rm gw}(\eta, k) = \frac{k^3}{48\pi^2}\sum_{\lambda = +, \times}\Big[P_{A, \lambda}(k) +  P_{B, \lambda} ( k) \Big].
\end{align}
Finally, in order to obtain the gravitational wave spectrum \textit{today}, it is possible to account for its dilution following the end of radiation domination as follows \cite{Kite:2021yoe}: 
\begin{equation}
     \Omega_{\rm gw}(k) = \Omega_{r}\frac{g_*(k)}{g_{\star, 0}}\Big(\frac{g_{s, 0}}{g_s(k)}\Big)^{4/3}\Omega_{\rm gw}^{(\rm RD)}(k),
     \label{eq_today}
\end{equation} 
which is essentially a consequence of conservation of entropy. In the rest of the work, we factor out this factor for clarity. 

We can already see the convenience of the oscillation-envelope split, which focuses by construction on the quantities to which observations are sensitive. In the following section we finally provide an efficient method with which the fields $A_{\lambda, \bm k}, B_{\lambda, \bm k}$ can be computed in practice, which is in fact the core result of this work.

We conclude this section by presenting the well known analytic expression for $\Omega_{\rm gw}^{(\rm RD)}$ in the case of Gaussian initial conditions (where the four-point function needed to compute the spectrum is obtained applying Wick's theorem):
\begin{align}
   & \Omega_{\rm gw}^{(\rm RD, Gauss)}(k) = \int_0^1 dq\int_1^\infty ds~T(q, s)\times\nonumber\\
    &\times\Delta_\zeta^2\Big(\frac{k}{2}(s + q)\Big)\Delta_\zeta^2\Big(\frac{k}{2}(s - q)\Big),
    \label{eq_semi_anal}
\end{align}
where the integral kernel $T(q, s)$ reads explicitly \cite{Witkowski:2022mtg}:
\begin{align}
    &T(q, s) = \frac{12(q^2 + s^2 - 6)^4}{(s^2 - q^2)^8}(q^2 - 1)^2(s^2 - 1)^2\times\\
    &\times\Big[\Big(\log\Bigg|\frac{3 - q^2}{3 - s^2}\Bigg| + \frac{2(s^2 - q^2)}{q^2 + s^2 - 6}\Big)^2 + \pi^2\Theta(s - \sqrt{3})\Big].\nonumber
\end{align}
We will adopt this formula to benchmark our numerical method, evaluating the integral in Eq. \ref{eq_semi_anal} numerically, remarking that it can be done straightforwardly adopting simple methods such as a trapezoidal scheme or Simpson method. 
\section{An efficient lattice evaluation of the amplitudes}
\label{sec_lattice}
\begin{figure*}
    \centering
    \includegraphics[width=1\linewidth]{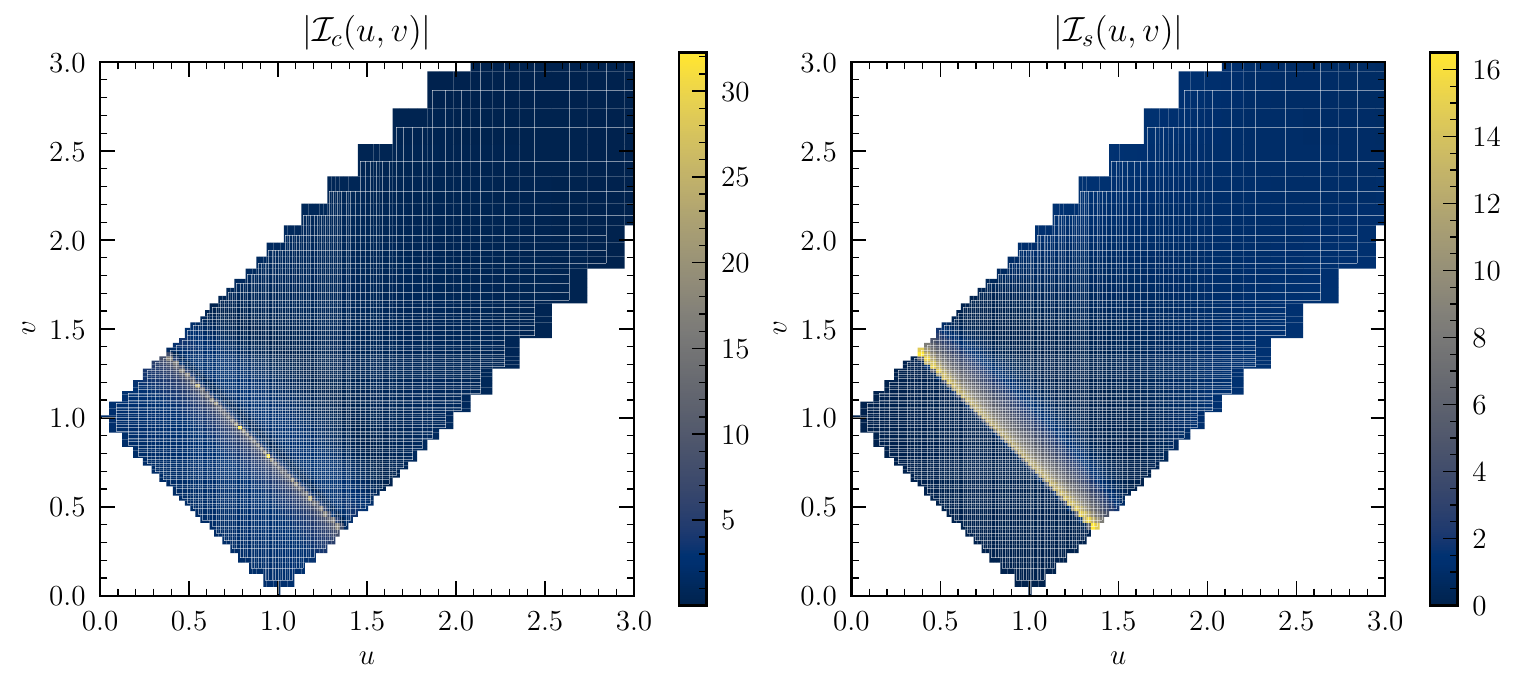}
    \caption{We show the relevant dynamical range of the kernels $\mathcal{I}_{m}$, restricting to the strip of the plane allowed by the condition $|u\pm v| <1$, arising because of the Dirac delta enforcing $\bm u + \bm v = \hat{\bm k}$ appearing in the Fourier integrals. We overlay the grid used to perform the decomposition into separable terms as explained in Sec. \ref{sec_decomposition}, noticing that it is focused where each kernel varies the most, becoming progressively more coarse at larger values of the arguments. We remark that because of its (integrable) logarithmic divergence, the peak value of $\mathcal{I}_c$ appearing in the plot depends on the discretization scale.}
    \label{fig_kernels}
\end{figure*}

The envelope equations can be formally integrated to obtain the asymptotic values to which the amplitudes settle for $\eta \gg 1/k$, defining the dimensionless variable $t = \eta k$, again focusing to the simple $w = 1/3$ case:
\begin{align}
     A_{\lambda, \bm k} &= -\frac{1}{k^2}\int_0^\infty dt~  t\sin t S_{\lambda,\bm k}(t)  ,\nonumber \\ \quad B_{\lambda, \bm k} &=  \frac{1}{k^2}\int_0^\infty dt~t\cos t S_{\lambda,\bm k}(t);
     \label{eq_time_integrals}
\end{align}
in Fourier space, the source is computed as follows:
\begin{align}
    \mathcal{S}_{\lambda, \bm k}(t)   = -4k^5e^{ij}_{\lambda}(\hat{\bm k})\int\frac{d^3\bm u}{(2\pi)^3}\int\frac{d^3\bm v}{(2\pi)^3}\times \nonumber \\\times (2\pi)^3\delta_D^{(3)}(\bm u + \bm v - \hat{\bm k })u_{i} v_{ j}\Phi_{k\bm u}\Phi_{k\bm v}f(t, u, v),
\end{align} 
in terms of the adimensional momenta $\bm u = \bm q_1/k$, $\bm v = \bm q_2/k$. The kernel $f(t, u, v)$ reads:
\begin{align}
    &f(t, u, v) = T(ut) T(vt)\nonumber  \\+ \frac{1}{2}\Big(T(ut) &+ t \partial_tT(ut)\Big)\Big(T(vt) +  t \partial_t T(vt)\Big).
\end{align}
therefore, inserting the expression of the source into the one of the asymptotic values of the envelopes, we obtain:
\begin{align}
     A_{\lambda, \bm k} = 4k^3e^{ij}_{\lambda}(\hat{\bm k}) \int\frac{d^3\bm u}{(2\pi)^3}\int\frac{d^3\bm v}{(2\pi)^3}\times\nonumber\\\times(2\pi)^3\delta_D^{(3)}(\bm u + \bm v - \hat{\bm k })u_{i} v_{ j}\Phi_{k\bm u}\Phi_{k\bm v}\mathcal{I}_s(u, v); 
     \label{eq_naive_A}
\end{align}
\begin{align}
     B_{\lambda, \bm k} =-4k^3e^{ij}_{\lambda}(\hat{\bm k}) \int\frac{d^3\bm u}{(2\pi)^3}\int\frac{d^3\bm v}{(2\pi)^3}\times\nonumber\\\times(2\pi)^3\delta_D^{(3)}(\bm u + \bm v - \hat{\bm k })u_{i} v_{ j}\Phi_{k\bm u}\Phi_{k\bm v}\mathcal{I}_c(u, v), 
     \label{eq_naive_B}
\end{align}
having defined the following symmetric kernels:
\begin{align}
    \mathcal{I}_s(u, v) &:= \int_0^\infty dt~  t\sin t f(t, u, v) , \nonumber\\ \mathcal{I}_c(u, v) &:= \int_0^\infty dt~  t\cos t f(t, u, v) ;
\end{align}
remarkably, these integrals can be solved analytically; we adopt the results of \cite{Kohri:2018awv}, paying attention to a relative factor due to different conventions:
\begin{align}
    \mathcal{I}_s(u, v)  &= \frac{27\pi (u^2 + v^2 - 3)^2}{32u^3v^3 }\Theta(u +v - \sqrt{3}), \quad \nonumber\\\mathcal{I}_c(u, v) &= \frac{27 (u^2 + v^2 - 3)}{32u^3v^3 }\Big(-4uv + \nonumber\\&+(u^2 + v^2 - 3)\log \Big|\frac{3 - (u + v)^2}{3 - (u -v)^2}\Big|\Big).
    \label{eq_kernels}
\end{align}
We show these kernels as heatmaps in Fig. \ref{fig_kernels}. 
While the method of variation of constants is a well-established technique, we propose to include it in a lattice simulation as a way to overcome the costs associated to a direct time integration of the wave equation, with the direct advantage of focusing on the asymptotic value of the envelopes $A$, $B$, without the need to explicitly account for fast oscillations, which are in any case of no observational relevance.
The result of Eq. \ref{eq_kernels} is perfectly fine as it is, provided that one is interested in semi-analytic methods involving local non-Gaussianities of the kind $\zeta = \zeta_g + F_{\rm NL}\zeta_g^2  + G_{\rm NL}\zeta_g^3 + \dots$, allowing to obtain a closed form of $\Delta^2_{h}$ in terms of a finite number of Wick's contractions. For the sake of a lattice computation, needed for instance if the non-Gaussianities under consideration don't admit a perturbative expansion, this direct approach presents a severe computational bottleneck. Given a lattice realization of $\Phi = 2\zeta/3$, the integrals connecting it to the asymptotic value of the envelope amplitudes $A, B$ are basically impossible to evaluate directly. In fact, the kernels $\mathcal I_{s, c}$ are not separable, in the sense that they cannot be written as $\mathcal{I}(u, v) = \phi(u) \phi(v)$, so that the Fourier integral cannot be expressed as a convolution. This prevents us from integrating numerically the above equations. In fact, a brute force integration $f_{\bm k}\sim \int_{\bm q}F_{\bm q, \bm k} g_{\bm q}g_{\bm k}$ has a $O(N^6)$ complexity, while a convolution $f_{\bm k}\sim \int_{\bm q} g_{\bm q}g_{\bm k - \bm q}$ can be computed very efficiently applying the convolution theorem and computing numerically the needed direct and inverse Fourier Transforms adopting the Fast Fourier Transform (FFT) \cite{10.5555/1074100.1074393}, whose complexity is $O(N^3\log N)$.
The simplest way to take advantage of this algorithmic simplification is to numerically integrate Eq. \ref{eq_time_integrals}, computing the source at each time by simply convolving the evolved potentials. The problem is that while $A$ and $B$ eventually settle to constant values, the source term oscillates and contains many time scales, all of which have to be resolved by the time step chosen in the numerical integration. For this reason, it is worth to investigate whether it may be possible to bypass the need to perform a numerical integration. While the resulting kernels $\mathcal{I}_c$, $\mathcal{I}_s$ posses no simple factorization, it may be possible to find a family of functions $\phi^{(m)}_\alpha$ and normalization factors $\sigma^{(m)}_\alpha$ $(m = c, s)$ such that:
\begin{equation}
    \mathcal{I}_{m}(u, v) =\sum_{\alpha} \sigma^{(m)}_\alpha\phi^{(m)}_\alpha(u)\phi^{(m)}_\alpha(v),
    \label{eq_decomposition}
\end{equation}
assuming to be able to approximate the original functions up to the wanted precision with a finite number of terms. For the moment, let us assume that such a decomposition is possible: we provide an explicit construction in Sec. \ref{sec_decomposition}. Under such hypothesis we can write Eq. \ref{eq_naive_A} as:
\begin{align}
     A_{\lambda, \bm k} &= 4k^3e^{ij}_{\lambda}(\hat{\bm k}) \sum_{\alpha} \sigma_\alpha\int\frac{d^3\bm u}{(2\pi)^3}\int\frac{d^3\bm v}{(2\pi)^3}(2\pi)^3\times\\\times &\delta_D^{(3)}(\bm u + \bm v - \hat{\bm k })\Big[u_i\phi^{(s)}_\alpha(u)\Phi_{k\bm u}\Big]\Big[v_j\phi^{(s)}_\alpha(v)\Phi_{k\bm v} \Big] \nonumber,
\end{align}
and similarly for $B_{\lambda, \bm k}$. With a few more computations we can now rewrite this double integral in a manageable form. In fact, we need to go back to the old dimensionful variables $\bm q_1 = k\bm u, \bm q_2 = k\bm v$:
\begin{align}
     &A_{\lambda, \bm k} = \frac{4 e^{ij}_{\lambda}(\hat{\bm k})}{k^2}\sum_{\alpha} \sigma_\alpha\int\frac{d^3\bm q_1}{(2\pi)^3}\int\frac{d^3\bm q_2}{(2\pi)^3}\times(2\pi)^3\times\\&  \times\delta_D^{(3)}(\bm q_1 + \bm q_2 - \bm k )\Big[q_{1,  i}\phi^{(s)}_\alpha\Big(\frac{q_1}{k}\Big)\Phi_{\bm q_1} \Big]\Big[ q_{2,  j} \phi^{(s)}_\alpha\Big(\frac{q_2}{k}\Big)\Phi_{\bm q_2} \Big]\nonumber,
\end{align}
Notice that the integral is scale dependent, in the sense that the auxiliary vectors $\bm q_j\phi_\alpha \Phi_{\bm q_j}$ depend on $k$ through $\phi_\alpha = \phi_\alpha(q_j/k)$. For this reason, this approach is not able to yield a complete realization of $A, B$ on the whole grid with a single convolution. However, we can focus on a specific $\bm k$ bin and repeat the computation for each of them. In practice, for a fixed $k$ we define:
\begin{equation}
    V^{(m)}_{\alpha, \bm q}(k) := \phi^{(m)}_\alpha\Big(\frac qk\Big)\Phi_{\bm q},
\end{equation}
in terms of which we can write
\begin{align}
     &A_{\lambda, \bm k} = \frac{4 e^{ij}_{\lambda}(\hat{\bm k})}{k^2}\sum_{\alpha} \sigma_\alpha\int\frac{d^3\bm q_1}{(2\pi)^3}\int\frac{d^3\bm q_2}{(2\pi)^3}\times(2\pi)^3\times\\&  \times\delta_D^{(3)}(\bm q_1 + \bm q_2 - \bm k )\Big[q_{1,  i}V^{(s)}_{\alpha, \bm q_1}(k)\Big]\Big[ q_{2,  j} V^{(s)}_{\alpha, \bm q_2}(k) \Big]\nonumber.
\end{align}
We can define the real-space counterparts $V_\alpha^{(m)}(k, \bm x)$:
\begin{equation}
    V_\alpha^{(m)}(k, \bm x) = \int \frac{d^3\bm q}{(2\pi)^3}e^{i\bm q \cdot \bm x}  V^{(m)}_{\alpha, \bm q}(k),
\end{equation}
in terms of which we can write the envelopes adopting the convolution theorem:
\begin{align}
     A_{\lambda, \bm k} &= -\frac{4 e^{ij}_{\lambda}(\hat{\bm k})}{k^2} \Big[\sum_{\alpha} \sigma_\alpha \partial_i V^{(s)}_\alpha(k, \bm x)\partial_j V_\alpha^{(s)}(k, \bm x)\Big]_{\bm k}, \nonumber\\\quad B_{\lambda, \bm k} &=\frac{4 e^{ij}_{\lambda}(\hat{\bm k})}{k^2} \Big[\sum_{\alpha} \sigma_\alpha \partial_i V_\alpha^{(c)}(k, \bm x)\partial_j V_\alpha^{(c)}(k, \bm x)   \Big]_{\bm k}.
     \label{eq_final_envelopes}
\end{align}
where the gradients $\partial_i V_\alpha^{(m)}(k, \bm x)$ in real space corresponds to the factors of $\bm q_{1}, \bm q_2$ in Fourier space:
\begin{equation}
    \partial_i V_\alpha^{(m)}(k, \bm x) = \int \frac{d^3\bm q}{(2\pi)^3}i q_ie^{i\bm q \cdot \bm x}  V^{(m)}_{\alpha, \bm q}(k).
\end{equation}
The power spectrum of a given field $X = \{A_\lambda, B_\lambda\}$, at a given wavenumber $k$, can be obtained adopting an ergodic approach, averaging the squared amplitude of the Fourier modes within a narrow shell centered around the mode of interest, i.e. for $k - \Delta k/2<|\bm k|<k + \Delta k/2$, with $\Delta k$ being the thickness of the shell:
\begin{equation}
    P_X(k) = \frac{1}{VN_{k}}\sum_{|\bm k| \in k \pm \Delta k/2} |X_{\bm k}|^2;
\end{equation}
where $N_k$ is the number of modes within the shell and we choose $\Delta k = \pi/L$, namely the fundamental mode.
We remark that even in the case where $\zeta$ (and hence $\Phi$) is Gaussian, the induced strain is not. Therefore, modes within the same shell in $\bm k$-space are not independent. In order to estimate the power spectrum at the mode $k$ and the associated error we employ the Jackknife resampling method \cite{Efron1982}. To do so, we first transform $X_{\bm k}$ to real space; the volume is then divided in $M^3$ sub-volumes. $M^3$ power spectrum measurements $\{P^{(i)}_X\}$ are obtained by leaving out one sub-volume at time. Finally, the Jackknife error is obtained as follows:
\begin{equation}
    \sigma^2_{P_X(k)} = \frac{M^3 - 1}{M^3}\sum_i (P^{(i)}_X(k) - \bar{P}_X(k))^2,
\end{equation}
where $\bar{P}_X$ is the average of the replica power spectra:
\begin{equation}
    \bar{P}_X(k) = \frac{1}{M^3}\sum_i P^{(i)}_X(k).
\end{equation}
In our simulations, we choose $M = 4$. The scale dependence of Eqs. \ref{eq_final_envelopes} could appear problematic, since it requires to repeat all the lattice computations for each modes of interest; however, the procedure is so efficient that this is not a problem at all, especially considered that measurements are sensitive to a fairly narrow frequency band, so that only few values of $k$ are needed. Furthermore, concentrating on a scale at a time allows us to use fairly small grids: in fact, the nonlinearities responsible for the generation of SIGW are fairly local in Fourier space, in the sense that vast majority of scalar modes contributing to a given tensor mode $k$ are concentrated around the resonance occurring at $\sim \sqrt{3} k$. For this reason, for every modes inside the band of observational interest $k\in (k_{\rm min}, k_{\rm max})$, the corresponding simulation box just needs to capture the relevant modes that give rise to $k$ and not the whole dynamical range $(k_{\rm min}, k_{\rm max})$, as is required in standard methods.

Eqs. \ref{eq_final_envelopes} represent our main result on the analytic side.
\section{Decomposing the Kernel as a Sum of Separable Functions}
\label{sec_decomposition}
In this section we explicitly construct a possible decomposition of the kernels $\mathcal{I}_m$, ($m = c, s$) in the form delineated in Eq. \ref{eq_decomposition}. Before diving deep into the details of the construction, we notice that the vectors $\bm u$, $\bm v$ are constrained to form a triangle together with $\hat{\bm k}$. This means that it is possible to limit the reconstruction of the kernels inside the region of the $(u, v)$ plane limited by the conditions $|u + v|<1$, $|u- v|<1$. We remark that reconstructing the kernels on the whole domain is not only unnecessary but computationally intractable. Indeed, in the allowed strip the kernels are quite ill-behaved; for instance, in the neighborhood of the origin, the kernel $\mathcal{I}_s$ diverges rapidly. Attempting to reconstruct such divergence would introduce numerical instabilities. For this reason, the regions outside of the strip defined by the conditions $|u \pm v|<1$ has to be excluded by hand since the beginning. 
\begin{figure}
    \centering
    \includegraphics[width=1\linewidth]{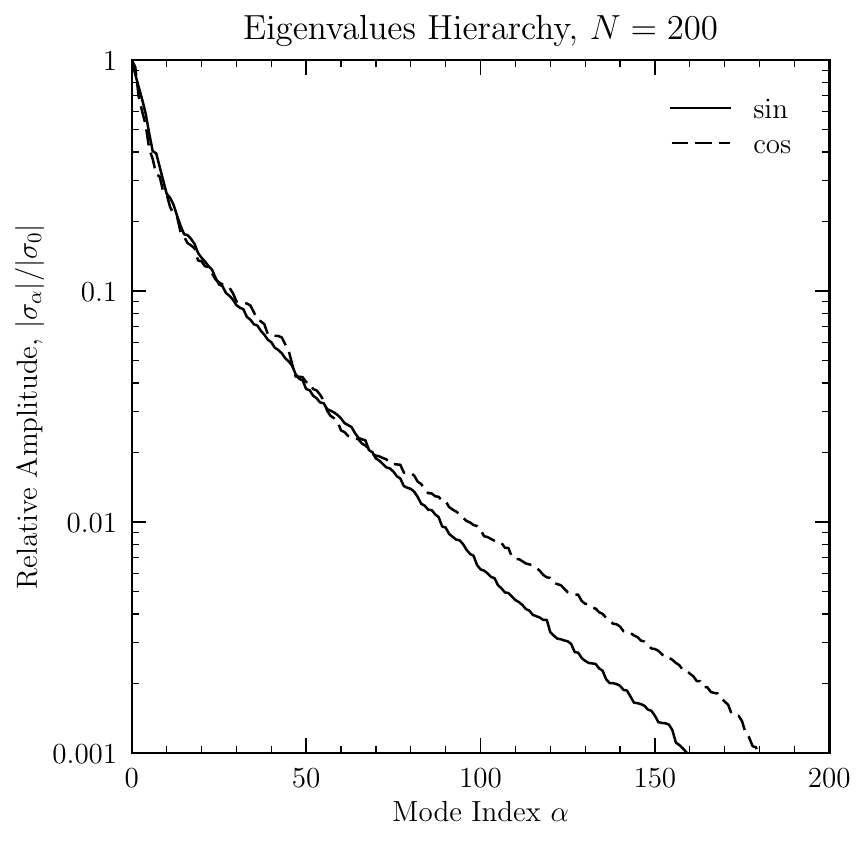}
    \caption{We show the absolute value of the eigenvalues (normalized to the largest one) relevant for the decomposition of the integral kernels $\mathcal{I}_s$, $\mathcal{I}_c$ (Eq. \ref{eq_diagonalization}), computed adopting the meta-parameters $u_{\rm min} = 10^{-3}$, $u_{\rm max} = 15$, $N_{\rm modes} = 200$, $u_{\rm p} = 1$, $\sigma_{\rm p} = 0.5$, $A_{\rm p} = 10$ and the grid-density function given by Eq. \ref{eq_grid_density}. The fall-off is very fast, and the inclusion of just $N_\alpha\sim 50-60$ modes is needed to reconstruct the kernel functions to a good degree.}
    \label{fig_eigenvalues}
\end{figure}
\begin{figure*}
    \centering
    \includegraphics[width=1\linewidth]{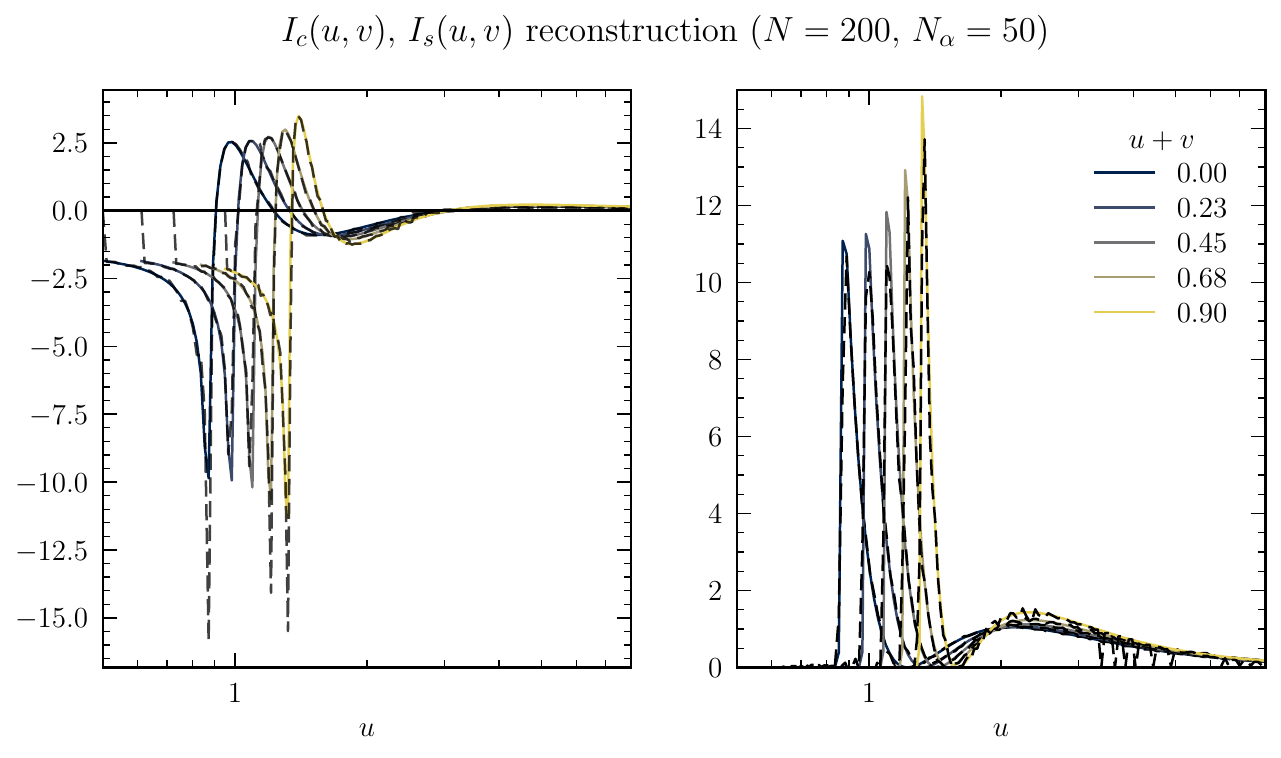}
    \caption{We compare the kernels with their reconstruction. We first adopted a grid built with $N = 200$ discretization points per side, covering the range $u_{\rm min} = 10^{-3}$, $u_{\rm max} = 15$, with a density of points given by Eq. \ref{eq_grid_density} and meta parameters $u_{\rm p} = 1$, $\sigma_{\rm p} = 0.5$, $A_{\rm p} = 10$. The kernels are reconstructed using only the first $N_\alpha = 50$ dominant modes in the eigenvalue decomposition of $C_{ij}$ (see Sec. \ref{sec_decomposition} for the details). The reconstruction is excellent in most of the dynamic range, apart on the edges ($u + v \approx 1$) for the sine kernel. However, since this function enters as an integral kernel, an error confined to a small part of the integration domain is forgivable.}
    \label{fig_reconstruction}
\end{figure*}
In the following we lighten the notation by dropping the denotation $m$. We start by considering a set of real basis functions $f_i$ so that:
\begin{equation}
    \mathcal{I}(u, v) = \sum_{i, j} C_{ij} f_i(u)f_j(v),
    \label{eq_decomposition_2}
\end{equation}
assuming them to be normalized so that it is possible to obtain the coefficients as follows:
\begin{equation}
    C_{ij} = \int du \int dv~f_i(u)f_j(v) \mathcal{I}(u, v).
\end{equation}
Once $C_{ij}$ has been computed, it can always be diagonalized since it is a real symmetric matrix. Let $q^{(\alpha)}_i$ be the $i$-th component of the $\alpha$-th eigenvector of $C_{ij}$, and $\sigma^{(\alpha)}$ the associated eigenvalue. We can therefore write:
\begin{equation}
    C_{ij} = \sum_\alpha \sigma_{\alpha} q_{i}^{(\alpha)}q_{j}^{(\alpha)};
    \label{eq_diagonalization}
\end{equation}
inserting this expression in Eq \ref{eq_decomposition_2}, we can rearrange the order of the sums to obtain:
\begin{equation}
    \mathcal{I}(u, v) =  \sum_\alpha \sigma_{\alpha} \sum_{i}q_{i}^{(\alpha)}f_i(u)\sum_{j} q_{j}^{(\alpha)} f_j(v),
\end{equation}
which can be brought in the form of Eq. \ref{eq_decomposition} by defining 
\begin{equation}
    \phi_\alpha (u) := \sum_{i}q_{i}^{(\alpha)}f_i(u).
\end{equation}
For instance, the basis could be taken to be a family of orthogonal polynomials (like Legendre, Laguerre, etcetera) or even plane waves. However, the kernels are quite ill-behaved functions, as they contain discontinuities and a logarithmic divergence. While these features pose no problem in a direct integration (like in the semi-analytic computation in the Gaussian case), they could lead to severe instabilities when expanded over a basis defined globally. For instance, if a Fourier expansion is chosen, the reconstruction of the Heaviside theta would present large oscillations at the boundary (the Gibbs phenomenon \cite{Gibbs1898-ur}). For these reasons, we opt for a localized basis. As a proof of principle, we choose each $f_i$ to be non-zero only over a given interval centered around $u_i$ and with width $L_i$: 
\begin{equation}
    f_i(u)  =\frac{\mathbbm{1}_i(u)}{\sqrt{L_i}},
    \label{eq_basis}
\end{equation}
where $\mathbbm{1}_{i}(u)$ is the indicator function of the $i$-th interval. We leave the study of more sophisticated choices (such as B-splines) to future works. Different intervals are taken to be not overlapping and to cover the relevant dynamical range of the kernel, namely where it is significantly different from zero. These functions form an orthonormal set: 
\begin{equation}
    \int dx ~ f_i(u) f_j(u) = \delta_{i j}.
\end{equation}
While being orthonormal, these functions forms a complete basis only approximately; in practice, they are nothing more than a discretized version of the position basis, which in the continuum limit is a continuously infinite family of Dirac deltas. The accuracy of such a reconstruction is however under control and can be systematically improved by choosing an appropriate grid, as we will see. 
Adopting the basis functions of Eq. \ref{eq_basis}, we then obtain:
\begin{align}
    C_{ij} = \int \frac{du}{\sqrt{L_i}} \int \frac{dv}{\sqrt{L_j}}~\mathbbm{1}_{i}(u)\mathbbm{1}_j(v) \mathcal{I}(u, v);
    \label{eq_C_matrix}
\end{align}
assuming to have chosen the intervals so that the kernel varies slowly within them, this reduces to 
\begin{equation}
    C_{ij}  \approx \sqrt{L_i L_j}\mathcal{I}(u_i, u_j).
    \label{eq_C_matrix_discrete}
\end{equation}
At face value, it could seem to be more convenient to adopt the exact form of Eq. \ref{eq_C_matrix}; however, this is inconsistent with the discretized basis used eventually to reconstruct the kernels, and in practice gives worse result than the adoption of Eq. \ref{eq_C_matrix_discrete}.

In order to improve the accuracy of this discretization procedure without using an unreasonable large number of discretization intervals, we can build the grid so to cover more finely the region of the plane where the kernels varies the most (see Fig. \ref{fig_kernels}). To do so, we build the one-dimensional grid so that the density of points is distributed according to a Gaussian centered on a given $u_{\rm pivot}$. We remark that the kernels vary mostly along the diagonal direction; therefore, it would be convenient, in principle, to adapt the grid to this feature. However, this is not possible; the 2D grid has to be a cartesian product of a discretization of the $u$-axis with itself in order for the decomposition to work. The reason is that a rotation of the coordinate $(u, v)$ would completely destroy the hope of separating the integrand of equations such as Eqs. \ref{eq_naive_A}, \ref{eq_naive_B}. 
We start by considering the (yet un-normalized) density of points to be uniform plus a bump centered over a pivotal point of interest $u_{\rm p}$: 
\begin{equation}
    \varrho(u) = \frac{dn}{du} = 1 + A_{\rm p}\exp\Big(-\frac{(u - u_{\rm p})^2}{2\sigma^2_{\rm p}} \Big);
    \label{eq_grid_density}
\end{equation}
written in terms of an amplitude $A_{\rm p}$ controlling the enhancement in density of sampling points around the pivot, and the relative width of said enhancement, $\sigma_{\rm p}$. $A_{\rm p} = 0$ simply leads to $\varrho = 1$, which corresponds to a uniform discretization of the $u-$grid. Let $N_{\rm modes}$ be the number of points in which we want to divide the range $[u_{\rm min}, u_{\rm max}]$; we can then find the normalized cumulative number of points: 
\begin{equation}
    n(u) = \frac{N_{\rm modes}}{\int_{ u_{\rm min}}^{u_{\rm max}}du'~\varrho(u')}\int_{ u_{\rm min}}^{u}du'~\varrho(u'),
\end{equation}
normalized so that $n(u_{\rm max}) = N_{\rm modes}$ indeed. Now, this function can be inverted; then the grid points computed as $u_n = u(n)$ for $n = 1, \dots, N_{\rm modes}$ are distributed according to the wanted density. In Fig. \ref{fig_kernels} we overlay to the heatmaps representing the kernels a grid constructed according to the algorithm we just presented, choosing $u_{\rm min} = 10^{-3}$, $u_{\rm max} = 15$, $N_{\rm modes} = 200$, $u_{\rm p} = 0.8$, $\sigma_{\rm p} = 0.5$, $A_{\rm p} = 10$. While this particular choice of meta-parameters is empirical, it is important to notice that the final result is not influenced by their precise value. In fact, a logarithmically spaced grid works as well, but it requires more discretization points. As a future improvement, it would be interesting to develop an algorithm able to find the optimal grid, namely the one able to minimize the reconstruction error for a given choice of $u_{\rm min}, u_{\rm max}, N_{\rm modes}$. For the time being we restrict to the analytic prescription of Eq. \ref{eq_grid_density} as a simple starting point, and we proceed to decompose the kernels. In Fig. \ref{fig_eigenvalues} we show the eigenvalues ordered according to their absolute value. Their quick fall-off renders useless the inclusion of modes beyond $N_\alpha\sim 50-60$. Notice that the kernels themselves are connected to the actual observables by an integral, which acts as a low-pass filter, desensitizing further the final result to the reconstruction errors. 
We show the kernels reconstructed by including only the first $N_\alpha = 50$ dominant eigenvalues in Fig. \ref{fig_reconstruction}. To plot each curve we fix a value of $u + v$ (the coordinate transverse to the diagonal in the $u-v$ plane) and plot the projection of the surface $\mathcal{I}_m(u, v)$ with respect to the $u$-direction. 
\section{Numerical Results}
\label{sec_numerical}

We implement numerically the algorithm described above using the Python library \texttt{PyTorch} \cite{paszke2019pytorchimperativestylehighperformance}, which enables to perform the computations both on a CPU and a GPU, the latter substantially accelerating the computation; the core choke-point of this algorithm is the necessity of performing many FFTs (one for each mode in which the kernels have been decomposed), which can be done in parallel with excellent performances on a GPUs, the only significant restriction being the memory needed to store the data. In all the benchmark we decompose the kernels over $200$ modes, including in the subsequent reconstruction the first $50$ dominant ones. Fixed the number of lattice spacing per side $N$, we determine the length of the simulation box by inverting the definition of the Nyquist frequency, $L = \pi N/k_{\rm max}$, itself chosen as
$k_{\rm max} = 3\max[k, \bar k]$. $\bar k$ is a wavenumber defined as follows:
\begin{equation}
    \frac{\int^{\bar{k}}_0d\log k ~\Delta_\zeta^2(k)}{\int^{\infty}_0d\log k ~\Delta_\zeta^2(k)} = 1 - \varepsilon,
\end{equation}
where $\varepsilon$ is a predetermined tolerance. In our simulations, we choose $\varepsilon = 0.1$. This definition of $k_{\rm max}$ allows to include enough power so to reconstruct both modes below the peak ($k\gtrsim k_*$, for which $\bar k$ is the relevant cutoff) and the ones around and above the peak, to which wavenumbers larger than $k$ itself do not contribute too much because of the sharp decline in the amplitude of the kernels after the resonant peak $u + v = \sqrt{3}$, translating into $q \approx \sqrt{3} k/2$. Notice that while in principle it would be tempting to further decrease $\varepsilon$, in practice it is not a good idea because that would lead to the exclusion of infrared modes if $N$ is not simultaneously increased. 

In all the following plots, the rescaling factor accounting for the dilution after the end of radiation era (Eq. \ref{eq_today}) is omitted for clarity. All the lattice computations are performed decomposing each kernel $N_{\rm modes} = 100$, using $N_\alpha = 50$ in the reconstruction step.
\subsection{Gaussian Initial Conditions}
\begin{figure}
    \centering
    \includegraphics[width=1\linewidth]{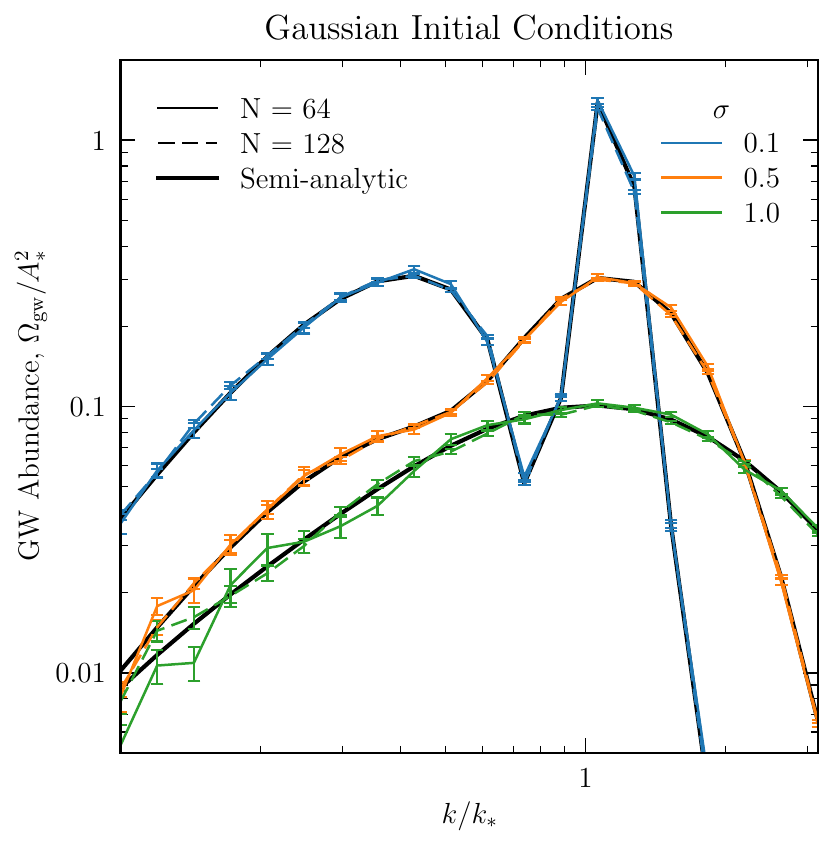}
    \caption{Gravitational wave spectrum induced by a Gaussian primordial curvature perturbation with a log-normal dimensionless power spectrum (Eq. \ref{eq_lognormal}), comparing the semi-analytic result (Eq. \ref{eq_semi_anal}, plotted as a continuous black line) with the outcome of lattice computations performed on a box with $N = 64$, $N = 128$ points per side (respectively continuous and dashed).}
    \label{fig_Gaussian}
\end{figure}
As a first test, we consider the case where $\Phi_{i, \bm k}$ is a Gaussian field, to be able to compare the outcome of the numerical simulation with a relatively simple semi-analytic expression. As a model for the small-scale power spectrum, we adopt the standard log-normal peak:
\begin{equation}
    \Delta_\zeta^2(k) = \frac{A_\star}{\sqrt{2\pi \sigma^2}}\exp\Big(-\frac{\log^2(k/k_\star)}{2\sigma^2}\Big)
    \label{eq_lognormal}
\end{equation}
showing that a sub $\sim 10\%$ percent accuracy is reached using a grid of just $N = 64$ per side, adopting $N_\alpha = 200$ for the decomposition of the kernel, keeping the $N_{\alpha}^{(\rm rec)} = 50$ with the largest eigenvalues for their subsequent reconstruction. In Fig. \ref{fig_Gaussian} we show the result for $\sigma = 0.1, 0.5, 1$ adopted simulation boxes with $N = 64$, $N =128$, comparing the result of each simulation to the analytic result of Eq. \ref{eq_semi_anal}.
\subsection{Local Non-Gaussianities}
\begin{figure*}
\includegraphics[width=0.5\linewidth]{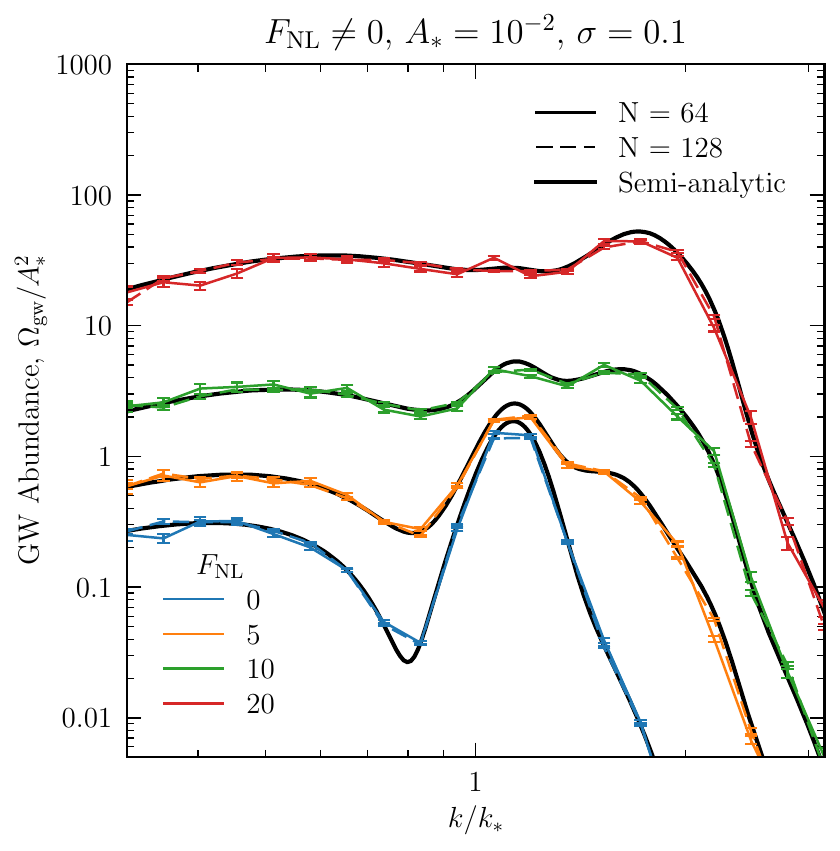}\includegraphics[width=0.5\linewidth]{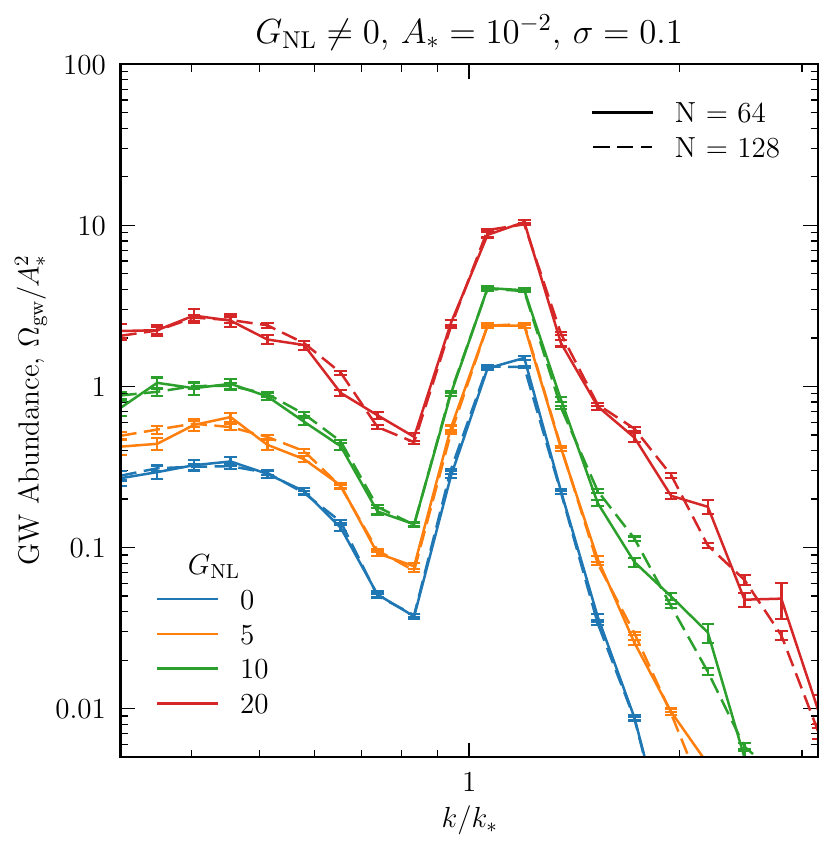}
    \caption{Gravitational wave spectrum induced by a non-Gaussian primordial curvature perturbation, expressed in terms of a Gaussian components by the local relation $\zeta = \zeta_g + F_{\rm NL}\zeta^2_g + G_{\rm NL} \zeta^3_g$; the Gaussian component is generated assuming the log-normal dimensionless power spectrum (Eq. \ref{eq_lognormal}) with $\sigma = 0.1$. We show the result of simulations performed adopting $N = 64$ (128) lattice points per sides as continuous (dashed) lines. In the case of $F_{\rm NL}\neq 0$ we compare them with the semi-analytic results of \cite{Perna:2024ehx} (credits to Gabriele Perna for the numerical data).}
    \label{fig_nonlocal}
\end{figure*}
\begin{figure*}
\includegraphics[width=0.5\linewidth]{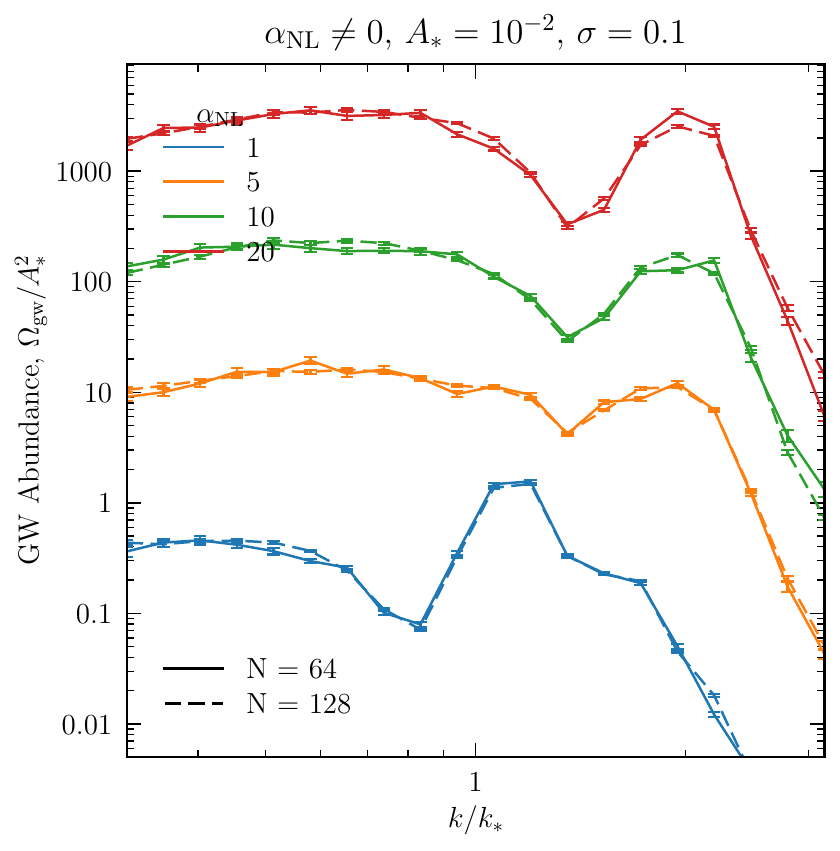}\includegraphics[width=0.5\linewidth]{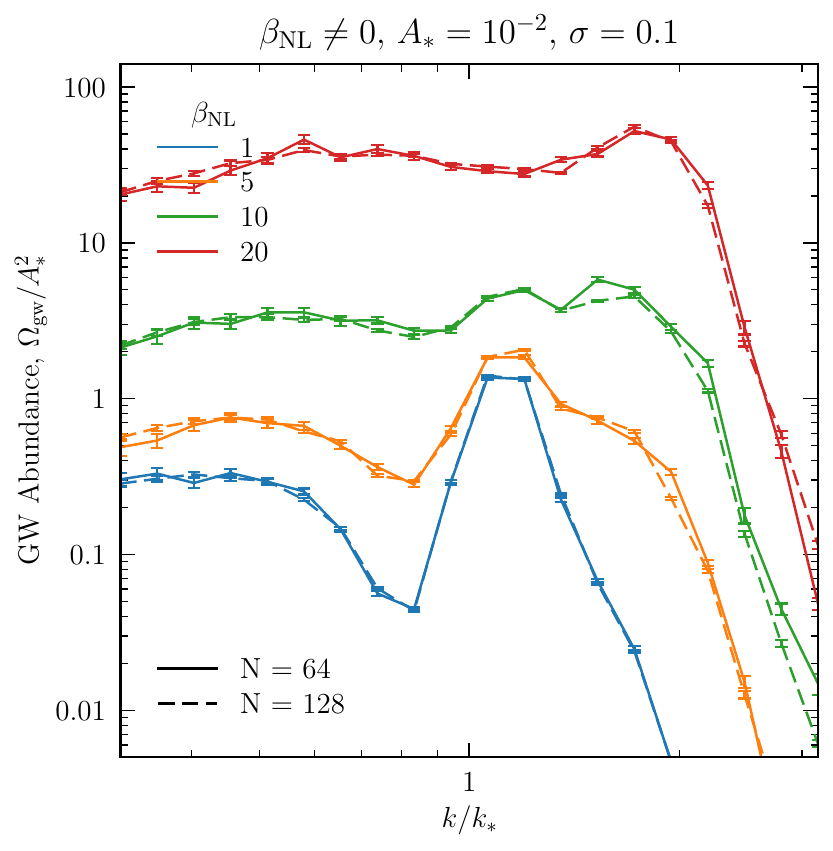}
    \caption{Gravitational wave spectrum induced by a non-Gaussian primordial curvature perturbation, expressed in terms of a Gaussian components in terms of the a perturbatively non-local relation of Eq. \ref{eq_nonlocal}; the Gaussian components is generated  assuming the log-normal dimensionless power spectrum (Eq. \ref{eq_lognormal}) with $\sigma = 0.1$. We compare the results of lattice computations performed on boxes with $N = 64$ and $N = 128$ point per side, respectively shown as continuous and dashed lines.}
    \label{fig_locnonGaussian}
\end{figure*}
We then move to the case where the primordial curvature perturbation is non-Gaussian, but can be expressed in terms of a Gaussian field $\zeta_g$ as follows:
\begin{equation}
    \zeta = \zeta_g + F_{\rm NL}\zeta^2_g  + G_{\rm NL} \zeta_g^3 + \dots,
\end{equation}
such an expansion can be obtained, for instance, from the $\delta N$ formalism; in this case, $\zeta_g = - \delta \phi/H\dot{\bar \phi}$, having decomposed the inflaton field as $\phi = \bar \phi + \delta \phi$. We generate the Gaussian field $\zeta_g$ adopting the log-normal peaked power spectrum (Eq. \ref{eq_lognormal}) as before. We consider separately the cases with $F_{\rm NL} = 1, 5, 10, 50$, $G_{\rm NL} = 0$, and $F_{\rm NL} = 0$, $G_{\rm NL} = 1, 5, 10, 50$, in both cases comparing the computations performed using $N = 64$, $N = 128$. The results are shown in Fig. \ref{fig_locnonGaussian}, compared with the outcome of the semi-analytic computations performed in \cite{Perna:2024ehx}, for which we thank the authors. We remark that these results are not normalization-independent anymore; because of the non-linear relation between $\zeta$ and $\zeta_g$, different $A_*$ distorts the overall shape. As known, a finite $F_{\rm NL}$ tends to flatten the peak structure present in the Gaussian case, while increasing the amplitude and the amount of power in the UV tail. For the pure $G_{\rm NL}$ case, it can instead be appreciated how for $G_{\rm NL}>1$ it holds an approximated scaling relation, $\Omega_{\rm gw}\sim G_{\rm NL}$, the shape being surprisingly similar across variations of $G_{\rm NL}$. 
\subsection{Non-Local Non-Gaussianities}
The $\delta N$ formalism is grounded on the separate Universe approximation, which has been shown to fail during a sudden transition between slow-roll and ultra-slow-roll phases of the inflationary evolution \cite{Jackson:2023obv, Briaud:2025ayt, Ahmadi:2026rzf}; in this circumstance, Laplacian corrections are relevant. Therefore, as a further application, we explore a simple model of perturbative non-Gaussianities beyond strict locality, considering the lowest possible non-Gaussian derivative terms: 
\begin{equation}
    \zeta = \zeta_g + \frac{\alpha_{\rm NL}}{k_*^2}\nabla^2\zeta_g^2 + \frac{\beta_{\rm NL}}{k_*^2} \zeta_g \nabla^2 \zeta_g + \dots,
    \label{eq_nonlocal}
\end{equation}
so that the value of $\zeta$ at a given point not only depends on the value of $\zeta_g$ at the same point but also on its immediate neighborhood. Such an expansion serves as a phenomenological approach to account for a derivative connection between the non-Gaussian curvature $\zeta$ and an unspecified Gaussian field $\zeta_g$. Once a specific model is established, its microphysics can be mapped onto the phenomenological coefficients $\alpha_{\rm NL}$ and $\beta_{\rm NL}$.
Here the scale $k_*$ is introduced so to deal with dimensionless constants, and in this study is taken to coincide with the location of the log-normal peak used to model the power spectrum of the Gaussian component. We show the results in Fig. \ref{fig_nonlocal}. We observe a good match between the results obtained with different grid resolutions. On a numerical level, this test is critical, as it proves the robustness of the code against the enhanced ultraviolet sensitivity introduced by the Laplacian operators.

\section{Conclusions}
\label{sec_conclusions}
Inflationary scenarios able to yield enhancements in the amplitude of primordial perturbations such as a transient phase of ultra-slow-roll evolution are characterized by strong non-Gaussianities, possibly scale-dependent, so that straightforward methods based on Wick's contractions may not be readily applicable. A convenient way to overcome such limitations are lattice simulations, based on a brute-force solution of the wave equation sourced by the effective stress-energy tensor sourced by nonlinearities in the Einstein equations, namely Eqs. \ref{eq_wave_tensor}, \ref{eq_source}. In this work, we presented an hybrid approach, combining a semi-analytic temporal integration (equivalent to an application of the Green's function method) with a fully non-perturbative lattice evaluation of the amplitudes of the induced strain. Such a method is very compelling since it completely circumvents the need of numerically integrating the wave equation on the lattice, focusing only on the observational relevant part of the strain, namely its slowly varying components. While straightforward on paper, this operation leads to a pair of expressions (Eqs. \ref{eq_naive_A} , \ref{eq_naive_B}) that are in practice impossible to compute, since they amount to computing for each point of the Fourier grid ($N^3$, if $N$ is the number of grid points per side) a Fourier integral that cannot be solved adopting FFT methods if left as it is. Thus, this apparent simplification results in a prohibitive complexity scaling of $O(N^6)$. The key point (which is the main contribution of this work) is that the original integral, by itself impossible to evaluate directly, can be decomposed into a rather small number ($N_\alpha \lesssim 60$) of convolutions, each efficiently computable using the convolution theorem and the FFT algorithm, bringing the complexity down to $O(N_\alpha N^3\log N)$. For $N \lesssim 150$ each FFT can be easily done even on a laptop, especially if GPU acceleration is available. For this reason, we implemented this method using the python library \texttt{PyTorch}, which allows to perform computations both on the CPU and the GPU.

We then benchmarked our results against the semi-analytic results valid in the case of Gaussian initial conditions and in presence of local non-Gaussianities, parametrized by $F_{\rm NL}$. The method proves to converge very quickly, being $N = 64$ already capable of yield an error consistently below $10\%$ with just $N_\alpha = 50$ modes used for the reconstruction of the kernel. We remark that increasing such number wouldn't lead to further improvements, a behavior consistent with the fall-off of the eigenvalues amplitudes as shown in Fig. \ref{fig_eigenvalues}.

We remark that these computations can be performed with modest computational resources. We performed the bulk of the development and the computations on a ScienceCloud virtual machine at the University of Zurich, equipped with a NVIDIA Tesla T4 GPU with $16$ GB of RAM. The resulting wall-clock time is particularly short: $1-2$ seconds for $20$ frequencies in the $N = 64$ case, $\sim 10$ seconds for $N = 128$. We therefore reckon this method to be a very useful tool to compute the power spectrum of SIGW in a fully non-perturbative fashion, including non-Gaussianities at all orders and overcoming the need of a direct numerical integration of the full wave equation.

It is necessary to point out that this work benefited from significant simplifications arising from the adoption of $w = 1/3$, such as the availability of simple analytical forms for the solution of the homogeneous wave equation, the transfer functions of the scalar potential, and ultimately the Fourier-space kernels. It is known that both in the PTA and LISA probe frequencies that entered the Horizon during a phase transition, respectively QCD and Electroweak; in both regimes the equation of state deviates from the simple case of $w = 1/3$, decreasing because of the change in number of effective degrees of freedom. For this reason, our results are intrinsically of limited immediate application to realistic modeling of observables pertaining to both bands. However, an extension of the method presented in this work to more realistic scenarios is in principle straightforward, apart from the complication of having to deal with numerical computations in order to obtain the quantities that were so far available in an analytic form. 

Another limitation of this work is the empirical choice of the meta-parameters defining the grid used to perform the kernel decomposition, namely $u_{\rm pivot}$, $A$, $N_\alpha$ etc. It would be interesting to find the optimal way to choose them in order to minimize the reconstruction error while keeping an economical number of modes, so to further decrease the computational weight. Such improvements are left to future works.

\begin{acknowledgments}
I warmly thank Gabriele Perna and Sabino Matarrese for discussions and early feedback. Moreover, I'm grateful to Gabriele Perna also for providing numerical values of the gravitational wave spectrum in the case of local non-Gaussianities.
Computations were in part performed on a ScienceCloud virtual machine at the University of Zurich, equipped with an NVIDIA Tesla T4 GPU with 16 GB of RAM. The python code \texttt{FLAN-SIGW}, together with notebooks with which the results and the plots presented in this paper can be reproduced, is made publicly available under MIT license on \href{https://github.com/giovannipiccoli99/FLAN-SIGW}{Github}. It is primarily built using the library \texttt{PyTorch} \cite{paszke2019pytorchimperativestylehighperformance}. The plots have been generated using \texttt{matplotlib} \cite{Hunter:2007} together with \texttt{smplotlib} \cite{https://doi.org/10.5281/zenodo.8126529}.
\end{acknowledgments}

\appendix


\bibliography{apssamp}

\end{document}